\documentclass[
    ,final            
  ]
  {aipproc}

\layoutstyle{6x9}

\usepackage{epsf,graphics,amssymb,amsfonts,amsmath,array}

\newcommand{\siml}{{\ \lower-1.2pt\vbox{\hbox{\rlap{$<$}\lower6pt\vbox{\hbox{$\sim$}}}}\ }} 
\newcommand{\simg}{{\ \lower-1.2pt\vbox{\hbox{\rlap{$>$}\lower6pt\vbox{\hbox{$\sim$}}}}\ }}

\newcommand{\sh}[1]{#1\hspace{-6pt}/}

\begin{document}

\title{Transverse momentum broadening and gauge invariance}

\classification{12.38.-t,12.38.Mh,13.87.-a}
\keywords      {Jet Quenching, SCET}

\author{Antonio Vairo}{
address={Physik-Department, Technische Universit\"at M\"unchen,
James-Franck-Str. 1, 85748 Garching, Germany}
}

\begin{abstract}
In the framework of the soft-collinear effective theory, we present a gauge invariant definition 
of the transverse momentum broadening probability of a highly-energetic collinear quark 
in a medium and consequently of the jet quenching parameter $\hat{q}$.
\end{abstract}

\maketitle

\section{Introduction}
Jet quenching occurs when in a heavy-ion collision 
an energetic parton propagating in one light-cone direction 
loses sufficient energy that few high momentum hadrons are seen in the final state,
where in the vacuum there would be a jet.
In this context, a parton is considered highly energetic when its momentum $Q$ is much larger 
than any other energy scale, including those characterizing the medium. 
Jet quenching, which has been observed at RHIC \cite{Adcox:2001jp} 
and at LHC \cite{Chatrchyan:2011sx}, manifests itself in many ways. 
In particular, the hard partons produced in the collision lose energy and 
change direction of their momenta. This last phenomenon goes under the name of 
transverse momentum broadening.

\begin{figure}[ht]
\makebox[-5truecm]{\phantom b}
\put(0,0){\epsfxsize=10truecm \epsfbox{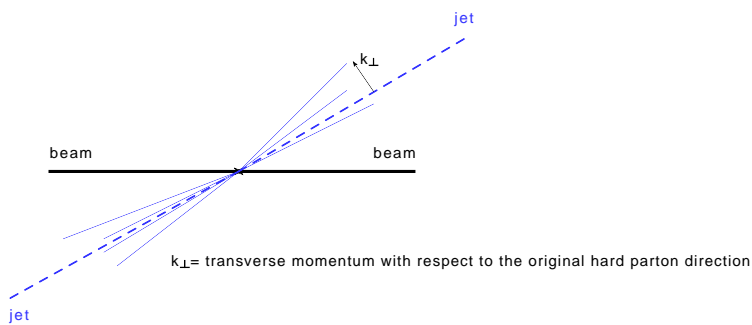}}
\caption{Kinematics of transverse momentum broadening.
\label{fig1}
}
\end{figure}

A way to describe the transverse momentum broadening is by means of the probability, $P(k_\perp)$, 
that after propagating through the medium for a distance $L$ ($\to \infty$) the hard parton acquires 
a transverse momentum $k_\perp$ (see Fig.~\ref{fig1}):
$\displaystyle \int \frac{d^2k_\perp}{(2\pi)^2}\, P(k_\perp) =1$.
A~related quantity is the jet quenching parameter, $\hat{q}$, which is  
the mean square transverse momentum picked up by the hard parton per unit distance traveled:
\begin{equation}
\hat{q} = \frac{1}{L} \int \frac{d^2k_\perp}{(2\pi)^2}\, k_\perp^2\, P(k_\perp)\,.
\label{qhat}
\end{equation}
In the following, we will review the derivation of a gauge invariant expression for $P(k_\perp)$ 
in the case of the propagation of a highly energetic quark.
The original detailed derivation can be found in \cite{Benzke:2012sz}.

\section{Scales and effective field theory}
\begin{figure}[ht]
\makebox[-4truecm]{\phantom b}
\put(0,0){\epsfxsize=8truecm \epsfbox{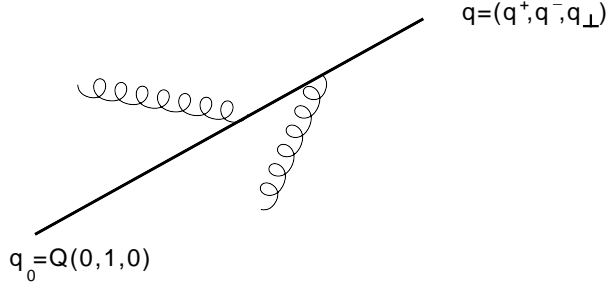}}
\caption{Parton momentum in light-cone coordinates.
\label{fig2}
}
\end{figure}

We consider a highly energetic quark of momentum $Q$ propagating along one light-cone direction $\bar{n}
= (1,0,0,-1)/\sqrt{2}$. The light-cone momentum coordinates are $q^+ = \bar{n}\cdot q$, 
$q^- = n\cdot q$, with $n= (1,0,0,1)/\sqrt{2}$, and $q_\perp$, which is the momentum component that is 
transverse with respect to the light-cone directions $n$ and $\bar{n}$, see Fig.~\ref{fig2}. 
If the quark propagates in a medium whose energy scales are much smaller than $Q$, 
then we can define a parameter $\lambda \ll 1$, which is the ratio of the 
energy scale characterizing the medium and $Q$. This small parameter may serve to classify 
the different modes of the propagating quark and interacting gluons.

\begin{figure}[ht]
\makebox[-6truecm]{\phantom b}
\put(0,0){\epsfxsize=12truecm \epsfbox{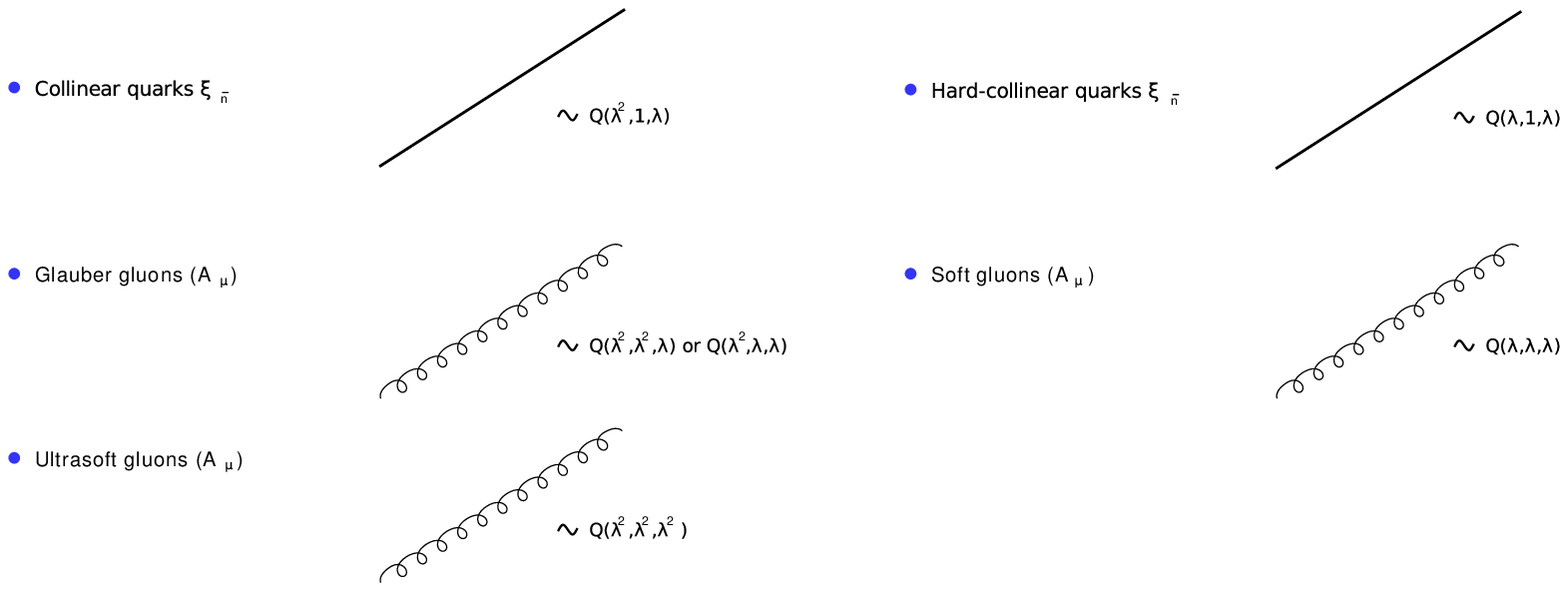}}
\makebox[0truecm]{\phantom b}\vspace{4mm}\\
\caption{Relevant degrees of freedom.
\label{fig3}
}
\end{figure}

We assume that the quark, after traveling along the medium, undergoes 
a transverse momentum broadening of order $Q\lambda$. If the virtuality of the 
quark is small, i.e. of order $Q^2\lambda^2$, then the parton has momentum 
$q \sim Q(\lambda^2,1,\lambda)$ and is called collinear.
We set up to describe the propagation of a single collinear 
quark in the medium.  A collinear quark may scatter in the medium 
with ultrasoft gluons, whose momenta scale like $Q(\lambda^2,\lambda^2,\lambda^2)$, 
with Glauber gluons , whose momenta scale like $Q(\lambda^2,\lambda,\lambda)$ or 
$Q(\lambda^2,\lambda^2,\lambda)$ or with soft gluons scaling like $Q(\lambda,\lambda,\lambda)$ 
through the emission of virtual hard-collinear quarks scaling like  $Q(\lambda,1,\lambda)$.
The relevant degrees of freedom are shown in Fig.~\ref{fig3}.

The effective field theory that describes the propagation of a collinear quark in the $\bar{n}$ light-cone direction 
is the soft-collinear effective theory (SCET) \cite{Bauer:2000ew} 
coupled to Glauber gluons \cite{Idilbi:2008vm}. After rescaling the quark field by 
$\displaystyle {\xi}_{\bar{n}} \to e^{-iQx^+} {\xi}_{\bar{n}}$,  
the Lagrangian may be organized as an expansion in $\lambda$:
\begin{equation}
\mathcal{L}_{\bar{n}}=\bar{\xi}_{\bar{n}}\,i\sh{n}\,\bar{n}\cdot D\,\xi_{\bar{n}}
+\bar{\xi}_{\bar{n}}\,\frac{D_\perp^2}{2Q}\,\sh{n}\,\,\xi_{\bar{n}}
+\bar{\xi}_{\bar{n}}\,i\frac{gF^{\mu\nu}_\perp}{4Q}\,\gamma_\mu\gamma_\nu\,\sh{n}\,\,\xi_{\bar{n}}\,
+ \dots \,,
\end{equation}
where $iD_\mu=i\partial_\mu+gA_\mu$ and $F^{\mu\nu}_\perp = i[D^{\mu}_\perp,D^{\nu}_\perp]/g$ is the  gluon field strength.
The fragmentation of the collinear quark into collinear 
partons is not taken into account by the above Lagrangian; 
a preliminary study of this effect can be found in \cite{D'Eramo:2011zz}.

\section{Momentum broadening in covariant gauges}
Collinear and hard-collinear quark fields, $\xi_{\bar{n}}(x)$, scale in the same way.
The operators $\bar{n}\cdot \partial$ and $\nabla_\perp$ 
scale like $Q\lambda^2$ and $Q\lambda$ respectively when acting on a collinear field $\xi_{\bar{n}}(x)$, 
and both scale like $Q\lambda$ when acting on a hard-collinear field $\xi_{\bar{n}}(x)$. 
Soft gluon fields scale like $Q\lambda$ and ultrasoft gluon fields scale like $Q\lambda^2$,  
for they are homogeneous in the soft and ultrasoft scale respectively. 
In contrast, the power counting of Glauber gluons depends on the gauge.
The equations of motion require $A^+(x)$ to scale like $Q\lambda^2$.
In a covariant gauge, if the gluon field is coupled to a homogeneous soft source, 
this also implies that $A_\perp(x) \sim Q\lambda^2$. 
The leading order Lagrangian in $\lambda$ is then
\begin{equation}
\mathcal{L}_{\bar{n}}=\bar{\xi}_{\bar{n}}\,i\sh{n}\,\bar{n}\cdot D\,\xi_{\bar{n}}
+\bar{\xi}_{\bar{n}}\,\frac{\nabla_\perp^2}{2Q}\,\sh{n}\,\,\xi_{\bar{n}}\,.
\end{equation}
Because ultrasoft gluons decouple at lowest order from collinear quarks trough the field redefinition
$\displaystyle \xi_{\bar{n}}(x) \to {\rm P}\,\exp\,\left[
ig \int_{-\infty}^{x^-}dy\, \bar{n}\cdot A_{\rm us}(x^+,y,x_\perp)\right]\xi_{\bar{n}}(x)$, 
where P stands for the path ordering operator, 
only one relevant vertex involving either Glauber or soft gluons has to be taken into account:\\
\begin{figure}[h]
\makebox[-3.5truecm]{\phantom b}
\put(0,0){\epsfxsize=3.5truecm \epsfbox{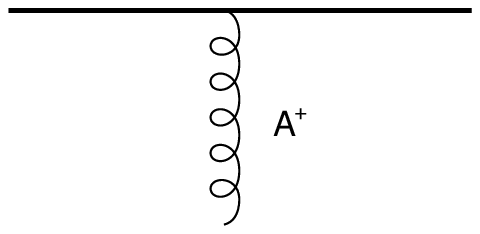}}
\put(115,25){$= igT^a\bar{n}^\mu \sh{n}$.}
\end{figure}

The transverse momentum broadening probability is then given by the imaginary part of the 
differential scattering amplitude 
\newpage
\begin{figure}[h]
\makebox[-5truecm]{\phantom b}
\put(0,0){\epsfxsize=9truecm \epsfbox{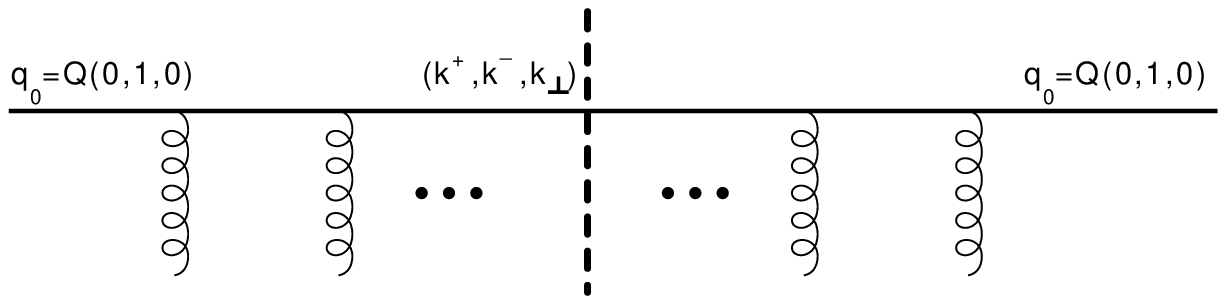}}
\put(275,30){$= P(k_\perp)\,,$}
\end{figure}
\makebox[0truecm]{\phantom b}\\
taken for $k_\perp \neq 0$ and normalized by the number of collinear quarks in the medium.
The scattering amplitude has the form (evaluated on a background of gluon fields)
$$
\int \!\prod_i \frac{d^4q_i}{(2\pi)^4} \cdots 
\frac{iQ}{2Qq_2^+-q_{2\perp}^2+i\epsilon}\sh{\bar{n}}\,
A^+(q_2-q_1)\sh{n}\,
\frac{iQ}{2Qq_1^+-q_{1\perp}^2+i\epsilon}\sh{\bar{n}}\,
A^+(q_1-q_0)\sh{n}\,
\xi_{\bar{n}}(q_0), 
$$
where the Dirac spinor $\xi_{\bar{n}}(q_0)$ satisfies $\sh{\bar{n}}\,\xi_{\bar{n}}(q_0) =0$ 
and is normalized as $\xi^\dagger_{\bar{n}}(q_0)\,\xi_{\bar{n}}(q_0) = \sqrt{2}Q$.
For Glauber gluons, the free propagator may be approximated by (e.g. in Feynman gauge)
\begin{equation}
D_{\mu\nu}(k) = D(k^2)g_{\mu\nu} \approx D(k^2_\perp)g_{\mu\nu}, 
\end{equation}
which implies that the scattering amplitude in coordinate space is at leading order 
\begin{equation}
\int dy^+d^2y_\perp\prod_i dy_i^- \,\cdots\,
\theta(y^-_3-y^-_2)A^+(y^+,y_2^-,y_\perp)\, \theta(y^-_2-y^-_1)A^+(y^+,y_1^-,y_\perp)\xi_{\bar{n}}(q_0)\,.
\label{wilexp}
\end{equation}
The same result also holds when considering the case of (hard-)collinear quarks interacting with soft gluons. 

\begin{figure}[ht]
\makebox[-4truecm]{\phantom b}
\put(0,0){\epsfxsize=7truecm \epsfbox{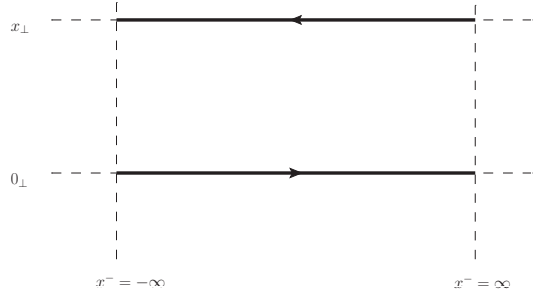}}
\caption{Wilson lines contributing to $P(k_\perp)$ in a covariant gauge.
\label{fig4}
}
\end{figure}

Because \eqref{wilexp} is just a term in the expansion of the Wilson line 
$\displaystyle W[y^+,y_\perp]$ $=$ $\displaystyle {\rm P}\,\exp\left[ig\int_{-L/\sqrt{2}}^{L/\sqrt{2}}\,dy^-A^+(y^+,y^-,y_\perp)\right]$ for $L\to\infty$, 
the transverse momentum broadening probability of a quark in covariant gauges is given by
\begin{equation}
P(k_\perp) = \int d^2x_\perp e^{ik_\perp\cdot x_\perp}\, 
\frac{1}{N_c} \left\langle {\rm Tr} \, \left\{ W^\dagger[0,x_\perp] W[0,0] \right\} \right\rangle\,,
\label{Pcov}
\end{equation}
where $\langle \dots \rangle$ denotes a field average. 
The Wilson lines of \eqref{Pcov} are shown in Fig.~\ref{fig4}. 
The relation between jet quenching and Wilson lines oriented along one of the light-cone directions was derived 
within different approaches in \cite{Baier:1996kr} and within SCET in \cite{D'Eramo:2010ak}.
Clearly the above expression is, in general, not gauge invariant (e.g. in the light-cone gauge $A^+=0$, one would have $W=1$).

\section{Momentum broadening in light-cone gauge}
In the light-cone gauge $A^+=0$, the free gluon propagator reads
\begin{equation}
D_{\mu\nu}(k) = D(k^2)\left( g_{\mu\nu}  - \frac{k_\mu\bar{n}_\nu + k_\nu \bar{n}_\mu}{[k^+]} \right)\,.
\end{equation}
For Glauber gluons $k_\perp/[k^+] \sim 1/\lambda$, which leads to on enhancement of order $\lambda$ in the 
singular part of the propagator. Moreover, because of the  $k_\perp/[k^+]$ singularity, one can write 
\cite{Belitsky:2002sm}
$A_\perp(x^+,x^-,x_\perp) = A^{\mathrm{cov}}_\perp(x^+,x^-,x_\perp)+A^{\mathrm{sin}}_\perp(x^+,x^-,x_\perp)$, 
where $A^{\mathrm{cov}}_\perp(x)$ contributes to the non-singular part of the propagator and 
vanishes at $x^-=\pm\infty$, while 
$A^{\mathrm{sin}}_\perp(x^+,x^-,x_\perp) = \theta(x^-)A_\perp(x^+,\infty,x_\perp)+\theta(-x^-)A_\perp(x^+,-\infty,x_\perp)$
with $A_\perp(x^+,\pm\infty,x_\perp) = -\nabla_\perp\phi^\pm(x^+,x_\perp)$.
The field $A_\perp$ does not vanish at infinity where it becomes pure gauge,
for the field tensor does (the energy of the gauge field is finite).

In the $A^+=0$ gauge, the scaling of the Glauber fields appearing in the Lagrangian changes to 
$A^{\rm cov}_\perp\sim Q \lambda^2$ and  $A^{\mathrm{sin}}_\perp\sim Q\lambda$.
The leading order Lagrangian in $\lambda$ is then   
\begin{equation}
\mathcal{L}_{\bar{n}}=\bar{\xi}_{\bar{n}}\,i\sh{n}\,\bar{n}\cdot \partial \,\xi_{\bar{n}}
+\bar{\xi}_{\bar{n}}\,\frac{(\nabla_\perp + ig A^{\mathrm{sin}}_\perp)^2}{2Q}\,\sh{n}\,\,\xi_{\bar{n}}\,,
\end{equation}
where gluons are just Glauber gluons. The relevant vertices are now two\\
\begin{figure}[h]
\makebox[-6truecm]{\phantom b}
\put(10,0){\epsfxsize=3.5truecm \epsfbox{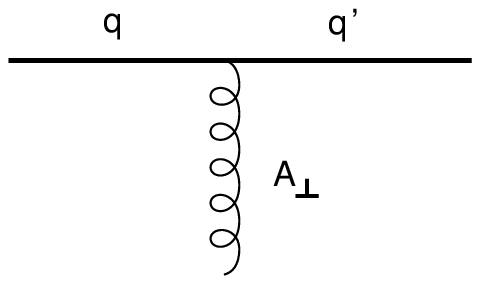}}
\put(125,25){$\displaystyle = -ig\,\frac{q'_\perp\cdot A^{\mathrm{sin}}_\perp(q'-q)+A^{\mathrm{sin}}_\perp(q'-q)\cdot q_\perp}{2Q}\,\sh{n},$}
\put(10,-80){\epsfxsize=3.5truecm \epsfbox{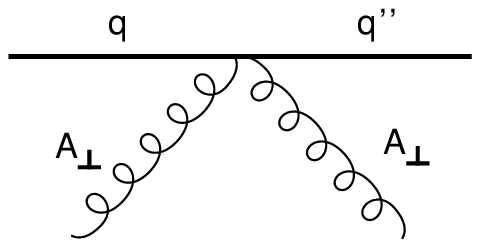}}
\put(125,-55){$\displaystyle = 
-\frac{ig^2}{2Q}\int\frac{\,d^4q'}{(2\pi)^4}\,A^{\mathrm{sin}\,i}_\perp(q''-q')A^{\mathrm{sin}\,i}_\perp(q'-q)\sh{n}.$}
\end{figure}
\makebox[0truecm]{\phantom b}\\
From the vertices one constructs the scattering amplitude (on the left of the cut)
\begin{figure}[h]
\makebox[-6truecm]{\phantom b}
\put(10,0){\epsfxsize=9truecm \epsfbox{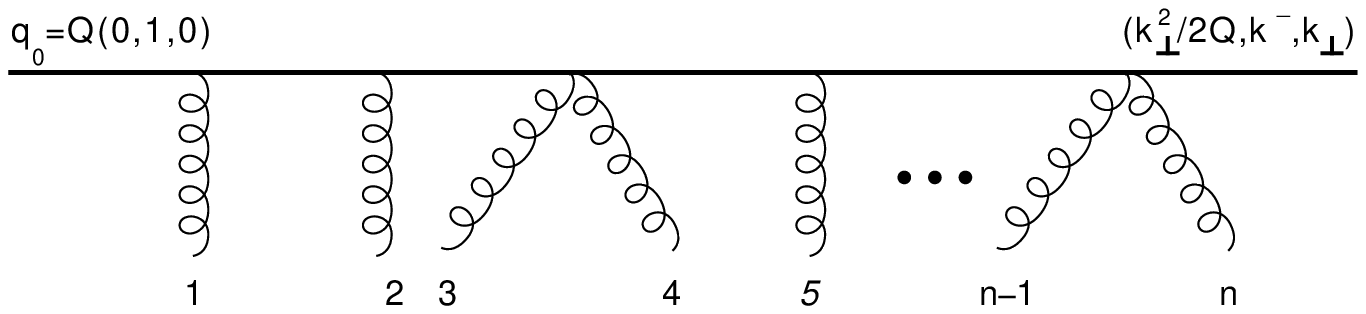}}
\put(270,25){$= G_n(k^-,k_\perp)\,.$}
\end{figure}
\makebox[0truecm]{\phantom b}\\
The function $G_n$ is a convolution of $G^+_{n-j}$, which involves only 
fields at $x^-=\infty$ and  $G^-_j$, which involves only fields at $x^-=-\infty$:
\begin{equation}
G_n(k^-,k_\perp)=
\sum_{j=0}^n\int\frac{\,d^4q}{(2\pi)^4}\,G^+_{n-j}(k^-,k_\perp ,q)\,
\frac{iQ\,\sh{\bar{n}}}{2Qq^+-q_\perp^2+i\epsilon}\,G^-_j(q)\,.
\end{equation}
The computation is done by solving recursively the equation (analogously for $G^+_n(q)$)
\begin{equation}
G^-_n(q)=
\int\frac{\,d^4q'}{(2\pi)^4}\,G^-_{n-1}(q')\raisebox{-15pt}{\includegraphics[width=20mm]{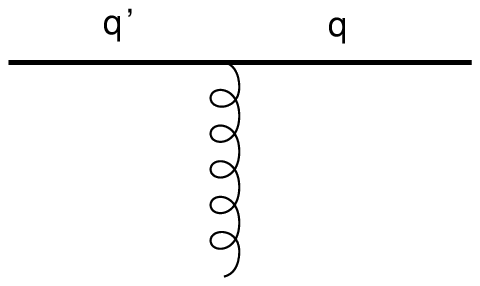}}
+\int\frac{\,d^4q''}{(2\pi)^4}\,G^-_{n-2}(q'')\raisebox{-10pt}{\includegraphics[width=20mm]{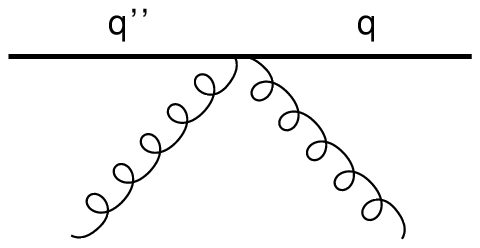}}\,,
\end{equation}
writing the differential amplitude as 
$$
\frac{1}{L^3\sqrt{2}Q}\int \frac{dk^+}{2\pi}\int \frac{dk^-}{2\pi} \; 2\pi \,Q\,  \delta(2Qk^+-k_\perp^2)
\; \bar{\xi}_{\bar{n}}(q_0)\, G_m^\dagger(k^-,k_\perp) \bar{\sh{n}}\,G_n(k^-,k_\perp)\,\xi_{\bar{n}}(q_0)\,,
$$
and eventually summing over all $m$ and $n$.
The expression of the transverse momentum broadening probability in light-cone gauge then reads
\begin{equation}
P(k_\perp) = \int d^2x_\perp e^{ik_\perp\cdot x_\perp}\, 
\frac{1}{N_c} \left\langle {\rm Tr} \, \left\{ 
T^\dagger[0,-\infty,x_\perp] T[0,\infty,x_\perp] 
T^\dagger[0,\infty,0] T[0,-\infty,0] 
\right\} \right\rangle \,,
\label{Plc}
\end{equation}
where $\displaystyle T[0,\pm\infty, x_\perp] = {\rm P}\, {\rm exp}
\left[\displaystyle  - ig\int_{-\infty}^{0} ds\, l_\perp \cdot A_\perp(0,\pm\infty,x_\perp+sl_\perp)\right]$ 
(for the definition of $T$ see also \cite{Idilbi:2010im}). The transverse vector $l_\perp$ is arbitrary.
The Wilson lines of \eqref{Plc} are shown in Fig.~\ref{fig5}. 

\begin{figure}[ht]
\makebox[-4truecm]{\phantom b}
\put(0,0){\epsfxsize=7truecm \epsfbox{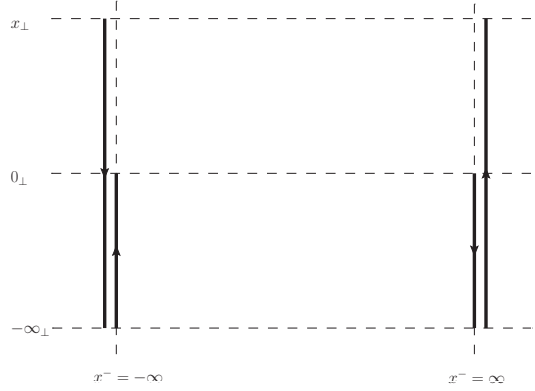}}
\caption{Wilson lines contributing to $P(k_\perp)$ in light-cone gauge. We have chosen $l_\perp \parallel x_\perp$.
\label{fig5}
}
\end{figure}

\section{Gauge invariant momentum broadening}
\begin{figure}[ht]
\makebox[-4truecm]{\phantom b}
\put(0,0){\epsfxsize=7truecm \epsfbox{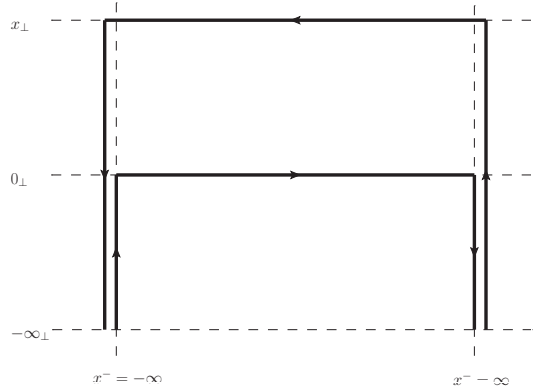}}
\caption{Wilson lines contributing to the gauge invariant expression of $P(k_\perp)$ given in \eqref{Pgi}. 
We have chosen $l_\perp \parallel x_\perp$. The fields at $x^-=\infty$ are contiguous while those at $x^-=-\infty$ are not. 
\label{fig6}
}
\end{figure}

Combining the results in covariant and light-cone gauge for $L\to\infty$, 
we obtain a gauge invariant expression for $P(k_\perp)$, which reads 
\begin{eqnarray}
P(k_\perp) \!\!&=& \!\! \int d^2x_\perp e^{ik_\perp\cdot x_\perp}\, 
\frac{1}{N_c} \left\langle {\rm Tr} \, \left\{ 
T^\dagger[0,-\infty,x_\perp] W^\dagger[0,x_\perp] T[0,\infty,x_\perp] \right.\right.
\nonumber\\
&& \hspace{4cm}
\times
\left.\left.
T^\dagger[0,\infty,0] W[0,0] T[0,-\infty,0] 
\right\} \right\rangle.
\label{Pgi}
\end{eqnarray}
The Wilson lines of \eqref{Pgi} are shown in Fig.~\ref{fig6}. Note that the fields are path ordered but not 
time ordered as in usual Wilson loops \cite{Brown:1979ya}. This difference should not be surprising 
since it reflects the fact that $P(k_\perp)$ describes the propagation of a single particle, 
while usual Wilson loops describe the propagation of a particle-antiparticle pair.

\begin{figure}[ht]
\makebox[-4truecm]{\phantom b}
\put(0,0){\epsfxsize=7truecm \epsfbox{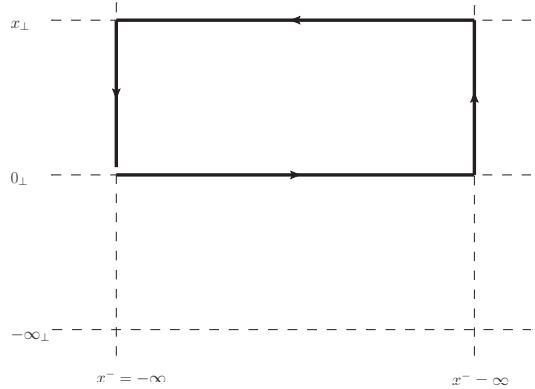}}
\caption{Wilson lines contributing to the gauge invariant expression of $P(k_\perp)$ given in \eqref{Pgi2}.
The fields in $(0,-\infty,0)$ are not contiguous.
\label{fig7}
}
\end{figure}

The expression of $P(k_\perp)$ may be simplified into 
\begin{equation}
P(k_\perp)=\int\,d^2x_\perp\, e^{ik_\perp\cdot x_\perp}
\frac{1}{N_c}
\left\langle\mathrm{Tr}\left\{[0,x_\perp]_-\, W^\dagger[0,x_\perp]\, [x_\perp,0]_+\, W[0,0] \right\}\right\rangle\,,
\label{Pgi2}
\end{equation}
where 
$\displaystyle [x_\perp,y_\perp]_\pm = {\rm P}\,\exp\left[-ig\int_{1}^0 ds\; (y_\perp-x_\perp)\cdot A_\perp(0,\pm\infty,x_\perp+s(y_\perp-x_\perp))\right]$, 
because contiguous adjoint lines cancel,
fields separated by space-like intervals commute and because of the cyclicity of the trace.
The Wilson lines of \eqref{Pgi2} are shown in Fig.~\ref{fig7}. 

The obtained expression for $P(k_\perp)$ does not depend on $l_\perp$. It is also gauge invariant.
In fact, under a gauge transformation $\Omega$,  
$\displaystyle {\rm Tr} \, \{ T^\dagger[0,-\infty,x_\perp] W^\dagger[0,x_\perp] \cdots T[0,-\infty,0] \}$
transforms to 
$\displaystyle   {\rm Tr} \, \{ \Omega[0,-\infty, -\infty\,l_\perp]
T^\dagger[0,-\infty,x_\perp] W^\dagger[0,x_\perp]$
$\displaystyle \cdots T[0,-\infty,0] 
\Omega^\dagger[0,-\infty, -\infty\,l_\perp] \}$, which is equal to the original expression, 
$\displaystyle {\rm Tr} \, \{ T^\dagger[0,-\infty,x_\perp] W^\dagger[0^+,x_\perp] \cdots T[0,-\infty,0] \}$,
after noticing that the fields in $\Omega[0,-\infty, -\infty\,l_\perp]$ commute with all the others 
(because of space-like separations) and after using the cyclicity of the trace.

\section{Conclusion}
Having derived the transverse momentum broadening probability, $P(k_\perp)$, we are in the position to write 
the jet quenching parameter $\hat{q}$ in a manifestly gauge invariant fashion:
\begin{eqnarray}
\hat{q} \!\! &=& \!\! \int\frac{d^2k_\perp}{(2\pi)^2}\,d^2x_\perp\,dx^- e^{i k_\perp\cdot x_\perp}
\frac{\sqrt{2}}{N_c}\Bigg\langle \mathrm{Tr}\Bigg\{
[0,x_\perp]_-U^\dagger_{x_\perp}[x^-,-\infty]\,gF_\perp^{+i}(0,x^-,x_\perp)\,
\nonumber\\
&&\hspace{18mm}
\times U^\dagger_{x_\perp}[\infty,x^-] [x_\perp,0]_+U_{0_\perp}[\infty,0]\,gF_\perp^{+i}(0,0,0)\,U_{0_\perp}[0,-\infty]
\Bigg\}\Bigg\rangle\,,
\label{qhat2}
\end{eqnarray}
where the fields $F_\perp^{+i}= \bar{n}\cdot\partial A_\perp^i-\nabla_\perp^iA^+ + ig[A^+,A_\perp^i]~$ 
come from the derivatives, $\nabla^i_\perp$, acting on the Wilson lines, 
and 
$\displaystyle U_{x_\perp}[x^-,y^-] = {\rm P}\,\exp\bigg[ig\int_{y^-}^{x^-}dz^- \,A^+(0,z^-,x_\perp)\bigg]$.
We recall that the above expression holds when the integral over $k_\perp$ has an ultraviolet cut-off of order $Q\lambda$, 
which is the size of the transverse momentum broadening that we have been considering.

\begin{theacknowledgments}
I thank Michael Benzke, Nora Brambilla and Miguel Angel Escobedo for collaboration on the 
work presented here. I acknowledge financial support from the DFG cluster of excellence 
``Origin and structure of the universe'' (http://www.universe-cluster.de). 
\end{theacknowledgments}

\bibliographystyle{aipprocl}

\end{document}